# Tuning SMSI Kinetics on Pt-loaded TiO$_2$(110) by Choosing the Pressure: A Combined UHV / Near-Ambient Pressure XPS Study

*Philip Petzoldt$^{‡1}$, Moritz Eder$^{‡1}$, Sonia Mackewicz$^1$, Monika Blum$^{2,3}$, Tim Kratky$^4$, Sebastian Günther$^4$, Martin Tschurl$^1$, Ueli Heiz$^{1}$\*, Barbara A. J. Lechner$^{5}$\**

$^1$Chair of Physical Chemistry

$^4$Physical Chemistry with Focus on Catalysis

$^5$Functional Nanomaterials

Department of Chemistry & Catalysis Research Center

Technical University of Munich, Lichtenbergstr. 4, 85748 Garching

$^2$Advanced Light Source, Lawrence Berkeley National Laboratory

1 Cyclotron Road, Berkeley, CA 94720, United States

$^3$Chemical Sciences Division, Lawrence Berkeley National Laboratory, 1 Cyclotron Road, Berkeley, CA 94720, United States

**Abstract**

Pt catalyst particles on reducible oxide supports often change their activity significantly at elevated temperatures due to the strong metal-support interaction (SMSI), which induces the formation of an encapsulation layer around the noble metal particles. However, the impact of oxidizing and reducing treatments at elevated pressures on this encapsulation layer remains controversial, partly due to the 'pressure gap' between surface science studies and applied catalysis. In the present work, we employ synchrotron-based near-ambient pressure X-ray photoelectron spectroscopy (NAP-XPS) to study the effect of $O_2$ and $H_2$ on the SMSI-state of well-defined $Pt/TiO_2(110)$ catalysts at pressures of up to 0.1 Torr. By tuning the $O_2$ pressure, we can either selectively oxidize the $TiO_2$ support or both the support and the Pt particles. Catalyzed by metallic Pt, the encapsulating oxide overlayer grows rapidly in $1\times10^{-5}$ Torr $O_2$, but orders of magnitudes less effective at higher $O_2$ pressures, where Pt is in an oxidic state. While the oxidation/reduction of Pt particles is reversible, they remain embedded in the support once encapsulation has occurred.

# 1. Introduction

Titania fulfills various functions in different fields of catalysis, including that of support material in thermal catalysis,[1] protective agent and light harvester in (photo)electrocatalysis,[2,3] or single-crystalline model sample in fundamental mechanistic studies.[4,5] Despite extensive titania-related research, important aspects in its surface chemistry still remain elusive, in particular at elevated pressures and temperatures. One of the most prominent examples is the so-called strong metal-support interaction (SMSI) between noble metals and $TiO_2$,[6] which has been shown to lead to an encapsulation of the metal particles by a thin sub-stoichiometric layer of $TiO_x$ upon annealing under reducing conditions.[6-9] From a technological point of view, the SMSI phenomenon can be both desired and unwanted. Particle encapsulation can have a beneficial effect on selectivity[10] and stability[11] of the catalyst, which has inspired several studies in thermal catalysis[12] and electrochemistry,[13,14] but it might equally lead to deactivation. In combustion catalysis, different strategies have been tested to prevent or reverse SMSI-related deactivation, e.g. by the addition of stabilizing compounds[15] or sputtering of the surface after its exposure to high temperatures.[16] While some authors claim that the encapsulation can also be reversed by $O_2$ annealing[6,17,18-20], others report an oxidation of the particles and/or thickening of the titania encapsulation layer under such conditions.[7,21-23]

In their pioneering work on SMSI, Tauster et al. reported that the change in chemisorption behavior of a reduced $Pd/TiO_2$ powder catalyst upon $O_2$ annealing at 773 K is completely reversible.[6] Similarly, different authors found restored $H_2$ chemisorption after annealing reduced $TiO_2$ decorated with noble metal powder catalysts at elevated $O_2$ pressures and temperature (>70 Torr $O_2$, > 573 K).[17] For encapsulated Pt supported on a $TiO_2$ thin film, Dwyer et al. observed significantly increased CO desorption from Pt in TPD experiments after annealing in ~$7.5 \times 10^{-7}$ Torr $O_2$ at 875 K for less than 1h.[19] Contrary to those studies, Pesty et al. found no evidence for a de-encapsulation on a reduced $Pt/TiO_2(110)$ catalyst upon annealing in $1 \times 10^{-6}$ Torr $O_2$ at up to 1000 K for 10 min in low energy ion scattering experiments, but proposed the formation of a new stoichiometric TiO$_2$ layer on Pt particles.[7] Naitabdi et al. observed a considerable encapsulation of Pt on a TiO$_2$(110) support in 0.75 Torr O$_2$ at 440 K[22] and the TiO$_x$ overlayer was described to grow significantly upon the exposure of O$_2$ (here at ~ 750 Torr) at elevated temperature (600 °C) in a recent investigation.[21] In those studies, different platinum loadings (from larger nanoparticles to sub-monolayer amounts of platinum atoms) and different TiO$_2$ support materials (from powders to single crystals) have been studied in pressure regimes spanning several orders of magnitude. Furthermore, many applied catalytic studies employ mixtures of anatase and rutile, whereas most surface science studies focus on rutile TiO$_2$ single crystals. Both the rutile and the anatase phase of titania seem to be capable of encapsulating noble metal particles, though some studies report quantitative differences in their susceptibility to SMSI.[24] This vast parameter range likely accounts for the apparent discrepancy in the reported results.

A comprehensive and fundamental understanding of the encapsulation process is indispensable to allow the tuning of the encapsulation according to the application. In the present work, we systematically investigate highly defined Pt/TiO$_2$ model systems with near-ambient pressure X-ray photoelectron spectroscopy (NAP-XPS) under different oxidizing and reducing conditions. We compare samples prepared by deposition of size-selected Pt$_{10}$ clusters with particles from Pt vapor deposition on a rutile TiO$_2$(110) single crystal, with loadings ranging from 6% to 84% relative to the amount of Ti$^{4+}$ surface atoms.[25] We observe similar trends independently of the method of deposition during the first reduction step suggesting a comparable SMSI state regardless of particle size and loading. Furthermore, we show that it is the O$_2$ partial pressure which determines the oxidation behavior of Pt/TiO$_2$ at elevated temperatures (i.e., 800 K) and greatly influences the effectiveness of the SMSI encapsulation process. We also find that the effects of oxidation are only partially reversible by a subsequent reduction in ultra-high vacuum (UHV) or H$_2$. Our data explain the apparent contradiction of reported results in the literature demonstrating the importance of reaction environment on the SMSI process.

## 2. Experimental Section

All XPS experiments were carried out at beamline 9.3.2 at the Advanced Light Source (ALS), Lawrence Berkeley National Laboratory.[26] Bulk-reduced $TiO_2(110)$ rutile single crystals (Surface-net GmbH) were prepared by multiple cycles of $Ar^+$ ion sputtering (20 min, 1 keV, ~ $5x10^{-6}$ Torr), annealing in $O_2$ (20 min, 800 K, $1.5x10^{-6}$ Torr $O_2$), and annealing in UHV (15 min, 800 K) in a UHV setup at Technical University of Munich. Sample cleanliness was confirmed by Auger electron spectroscopy (AES) and reproducible thermal and photochemical reactivity checked by methanol temperature programmed desorption and photocatalytic conversion.[27] Two types of $Pt/TiO_2(110)$ samples were used, where Pt was deposited *ex situ* in the form of soft-landed $Pt_{10}$ clusters and *in situ* as evaporated atoms ('$Pt_{at}$'), respectively. For Pt clusters, our XPS measurements indicate the presence of sub-nm particles until the first annealing step ,in agreement with the literature.[20, 22] In the case of $Pt_{at}$, small particles of variable size form upon deposition.[28, 29] In both cases, Pt sinters to larger entities when heated to 800 K.[28, 29, 30] For details on the evolution of particle sizes upon annealing see section 4.1.

$Pt_{10}$ clusters were generated using a laser ablation cluster source and deposited under soft-landing conditions ($E_{kin}$ < 1 eV/atom) on the $TiO_2(110)$ support.[31] Pt was evaporated from a rotating metal target (99.95% purity, ESG Edelmetalle, Germany) by a frequency-doubled Nd:YAG laser (532 nm, 100 Hz, Spitlight DPSS, Innolas) and clusters were condensed by cooling the generated plasma in a supersonically expanding He pulse (He 6.0, Air Westfalen). Cationic clusters were guided, mass-selected in a quadrupole mass filter (Extrel, USA) and soft-landed onto the $TiO_2$ single crystal. The cluster loading was controlled by integrating the cluster neutralization current during the deposition. The cluster coverage of the $Pt_{10}$ sample reported here was 1.8% of a monolayer (ML) with respect to the amount of $Ti^{4+}$ surface atoms ($1.56x10^{15}$ atoms cm$^{-2}$). With an accessible crystal surface area of ~ 0.28 cm$^2$ in our setup, a cluster loading of 1.8% ML Pt$_{10}$, corresponds to ~8x10$^{13}$ Pt atoms. Employing the corresponding atom coverage of 18% ML, we will herein refer to the Pt cluster sample as Pt$_{10\_18\% \, ML}$. Pt cluster loaded samples were shipped and transferred to the ALS beamline in inert gas atmosphere (Ar and N$_2$) using a glove bag and glove box.

Pt$_{at}$ samples were prepared by e-beam evaporation of Pt atoms *in situ* at the ALS using a Pt rod (Goodfellow, 99.95%). Prior to the evaporation of Pt, the TiO$_2$(110) crystals were cleaned by another cycle of annealing in O$_2$ (here 0.1 Torr), Ar$^+$ sputtering, and annealing in UHV. Pt$_{at}$ loadings of 84% ML, 71% ML and 6% ML were achieved by varying the evaporation time, determined by comparison of the Pt 4f peak area with the Pt$_{10\_18\% \, ML}$/TiO$_2$ sample. In the following, these samples will be referred to as Pt$_{at\_84\% \, ML}$, Pt$_{at\_71\% \, ML}$, and Pt$_{at\_6\% \, ML}$. Scheme 1 gives an overview over all four samples and the respective treatments investigated in this work.

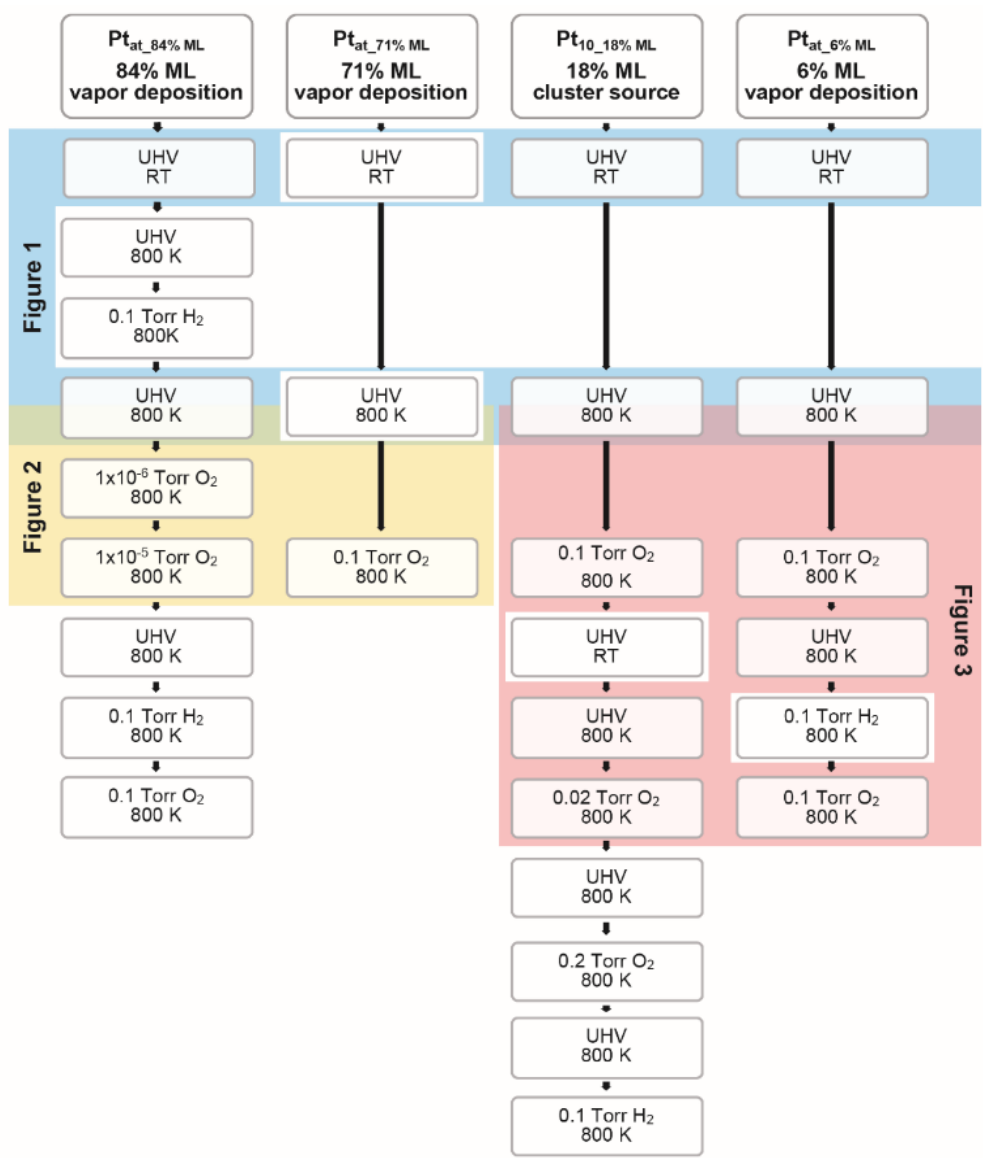

**Scheme 1.** Overview of all samples, i.e. Pt$_{at\_84\% ML}$, Pt$_{at\_71\% ML}$, Pt$_{10\_18\% ML}$, and Pt$_{at\_6\% ML}$, and the respective treatments investigated in this work where the subscript 'at' indicates atom deposition by *in situ* physical vapor deposition at the ALS experimental station and '10' indicates a sample with Pt$_{10}$ clusters deposited *ex situ* in our home laboratory at the Technical University of Munich. The abbreviation ML indicates the loading of total deposited Pt atoms with respect to the amount of Ti$^{4+}$ atoms on the TiO$_2$(110) surface in monolayers. Measurements corresponding to the Pt 4f spectra displayed in Figure 1, 2, and 3 are highlighted in blue, yellow, and red, respectively.

Every sample was characterized by a survey scan at the beginning of the XPS experiments (Figure S1). Carbon contaminations observed after Pt deposition were reliably removed in a first UHV annealing step (or 0.1 Torr $O_2$ annealing in case of $Pt_{10\_18\% ML}$; Figure S2). All experiments were conducted with an incident angle for X-ray photons of 15° and the electron analyzer placed perpendicular to the sample surface. The energy resolution of the beamline is ~$E/\Delta E=3000$.[26] Spectra were recorded using a VG-Scienta R4000 HiPP analyzer. $O_2$ (5.0, Praxair, USA) and $H_2$ (4.5, Praxair, USA) were dosed by backfilling the chamber, while the samples were heated by a pyrolytic boron nitride or an aluminum oxide button heater. For the photon energies employed here, beam-induced peak shifts could be excluded (Figure S3).

The Ti 2p peak position of $TiO_2$(110) at a binding energy of 458.5 eV[32] was used as energy reference for all given binding energies and especially for the one of the Pt 4f core level, since its energy position may be influenced by particle size effects. The chosen referencing procedure proved to be consistent, since the energy difference between the Ti $2p_{3/2}$ peak and the Ti 3s peak remained constant at a value of 396.38±0.07 eV (averaged over all samples) throughout all experiments and the derived binding energy of the Ti 3s peak amounted to 62.1 eV in agreement with values reported in the literature.[32, 33] In addition, a Au foil mounted in ohmic electrical contact with the samples was used as external reference in order to account for potential sample charging.

All spectra were normalized to the background level on the low binding energy side of the respective Pt 4f peak. Pt 4f core level peaks were fitted using Doniach-Sunjic (DS) functions convoluted with a Gaussian line shape. The Ti 3s plasmon is situated in close vicinity of the Pt 4f peak and had to be addressed by fitting a Voigt function (see below). The fitting also took into account a linear background which was separately determined by the overall slope of the photoelectron spectrum on the low binding energy side of the Pt 4f peak. For the $Pt_{at\_71\% ML}/TiO_2$ sample, a second Voigt function had to be added to account for an additional feature at a binding energy of ~ 74 eV, which is likely related to an Al contamination (Figure S4). Values for the branching ratio between Pt $4f_{7/2}$ and $4f_{5/2}$ peaks (1.33), the spin-orbit splitting (3.33 eV), and the Lorentzian width (0.32 eV FWHM) were fixed in accordance with literature values.[32, 34, 35] The Gaussian width and the asymmetry factor were first fitted without constraints and then averaged for each identified Pt compound. As a result, the following binding energies of the Pt $4f_{7/2}$ core level were extracted: 71.1-71.6 eV for as-deposited Pt particles, 71.2-72.2 eV for the $Pt_{mod}$ species, 70.9±0.03 eV for vacuum annealed samples with the Pt 4f peak shape of metallic Pt, and finally 72.8±0.2 eV and 74.60±0.3 eV for the two oxidic Pt species that occurred after oxidative treatment at elevated pressure. The average Gaussian width for all Pt 4f fits is 1.3 eV. For the Ti 3s plasmon, the Lorentzian and Gaussian width were determined from the Ti 3s peak first fitted without constraints and averaged for each sample to determine optimized parameters for a unique peak shape. The energy difference and intensity ratio between the Ti 3s peak and its plasmon peak were derived from a bare $TiO_2$(110) crystal. By employing these relations, the plasmon peak was fitted individually in every Pt 4f / Ti 3s spectrum based on the intensity of the Ti 3s peak. On average, the energy difference was determined to be 13.4±0.2 eV (Figure S4). After this, a refined, final fitting of all Pt 4f spectra was then conducted by applying optimized parameters with the $Pt_{mod}$ feature fitted with the same parameters as the metallic Pt 4f species. The final fit parameters for all species are listed in the supporting information (Table S1).

## 3. Results

The results from three different types of sample treatment are compiled in this section. First, we followed the changes caused by an initial reductive heating to 800 K in UHV. Second, we investigated the influence of oxygen exposure at different partial pressures at 800 K and, last, we monitored whether or not reductive heating in $H_2$ or UHV followed by $O_2$ exposure at various partial pressures leads to reversible or irreversible surface changes.

### *3.1 Pt 4f peak changes induced by reductive heating*

Figure 1 displays Pt 4f / Ti 3s spectra (55-84 eV) recorded for $Pt_{at\_84\%\ ML}$, $Pt_{at\_6\%\ ML}$, and $Pt_{10\_18\%\ ML}$ under UHV conditions at room temperature and at 800 K. Spectra obtained for $Pt_{at\_71\%\ ML}$ (Figure S5) under these conditions closely resemble those for $Pt_{at\_84\%\ ML}$ and are thus omitted from this comparison.

Prior to annealing, Pt/TiO$_2$ samples display a single Pt 4f peak (green) positioned at 71.2±0.1 eV (71.6 eV for $Pt_{at\_6\%\ ML}$ (Figure 1c)). The core level shifts and their dependency on the loading/particle size are in agreement with the literature for Pt supported on TiO$_2$(110).[36, 37] When held at 800 K in UHV, the Pt 4f peaks shift to lower binding energies for all samples until they reach 70.9 eV for $Pt_{at\_84\%\ ML}$ and $Pt_{10\_18\%\ ML}$ and 71.2 eV for $Pt_{at\_6\%\ ML}$, respectively. A binding energy of 70.9 eV is characteristic for the Pt $4f_{7/2}$ bulk core level of metallic Pt[35, 38, 39], indicating significant ripening of the clusters into large nanoparticles or extended Pt islands as expected at high temperatures.[28]

Apart from sintering, high-temperature reduction of Pt/TiO$_2$ catalysts is generally agreed to induce the formation of a thin sub-stoichiometric TiO$_x$ layer over Pt particles.[6, 7, 9] A slight decrease of the Pt 4f signal intensity and a narrowing of the peak width upon annealing observed for all samples investigated herein correspond to XPS spectra of encapsulated Pt particles[7, 40] and support the presumption of similar behavior in our experiments. This is further corroborated by a Pt 4f core level shift of ~ 0.2 eV to higher binding energies found for $Pt_{at\_84\%\ ML}$ upon prolonged annealing at 800 K (Figure S6), which has likewise been associated with the formation of a TiO$_x$ overlayer.[7] Annealing in 0.1 Torr H$_2$ does not have a significant effect on the Pt 4f peak compared to UHV annealing.

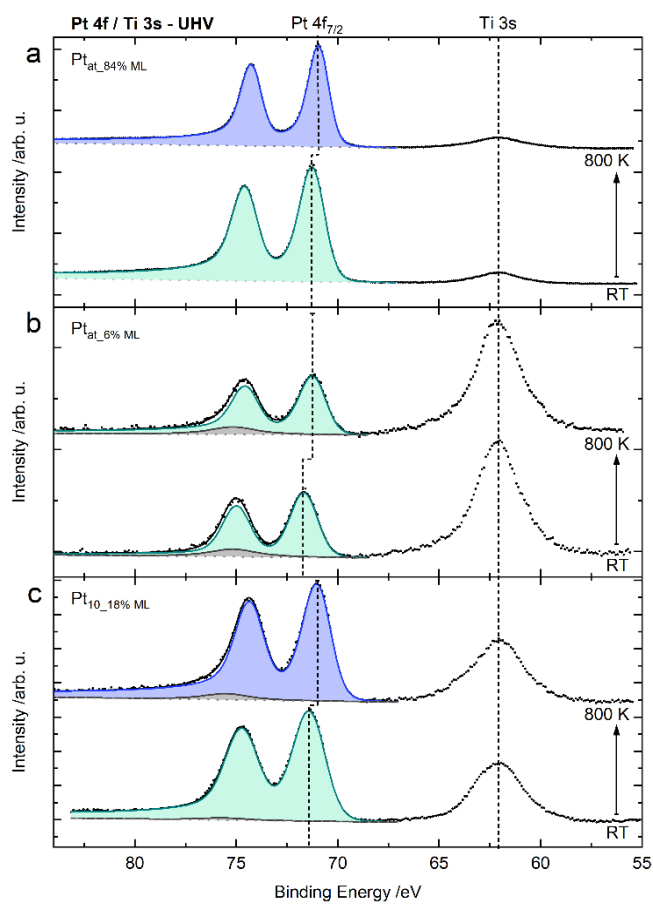

**Figure 1.** Pt 4f / Ti 3s spectra of Pt$_{at\_84\% ML}$ (a), Pt$_{at\_6\% ML}$ (b), and Pt$_{10\_18\% ML}$ (c) on TiO$_2$(110) measured in UHV at room temperature and at 800 K, recorded with an incident beam energy of 620 eV. Data are shown as black dots, fitted peak shapes as solid lines. Two Pt compounds can be identified: Pt particles at binding energies ~71.2-71.6 eV (shown in green; BE varies somewhat dependent on particle size) and bulk Pt at 70.9 eV (blue). Additionally, the Ti 3s plasmon (light gray) was fitted based on the intensity of the Ti 3s peak. Dashed lines indicate the position of the Pt 4f$_{7/2}$ and Ti 3s peak maxima. Spectra are offset for clarity.

*3.2 The Pt 4f core level peak during O$_2$ exposure at various partial pressures*

In order to elucidate the effect of oxygen annealing and potential pressure dependencies on the Pt/TiO$_2$ system, we measured Pt 4f / Ti 3s spectra at different O$_2$ pressures. Figure 2 displays the evolution of the Pt 4f signal for two samples with comparable Pt loadings ((a) Pt$_{at\_84\%\,ML}$ and (b) Pt$_{at\_71\%\,ML}$) at intermediate (1x10$^{-6}$ Torr and 1x10$^{-5}$ Torr) and high O$_2$ pressure (0.1 Torr), respectively. The applied O$_2$ exposure amounted to a dosage of ~ 3x10$^3$ Langmuir at 1x10$^{-6}$ Torr, ~ 3x10$^4$ Langmuir at 1x10$^{-5}$ Torr, and ~ 4x10$^8$ Langmuir at 0.1 Torr.

While the Pt 4f peak shape and binding energy are unaffected by annealing in 1x10$^{-6}$ Torr and 1x10$^{-5}$ Torr O$_2$ (Figure 2a, Figure S7), they change significantly in 0.1 Torr O$_2$ (Figure 2b, Figure S8). Almost immediately after applying the O$_2$ pressure, the main Pt 4f peak shifts by more than 0.5 eV toward higher binding energies. Moreover, its appearance loses the characteristic duplet of the metallic state and develops further components at higher binding energies. After ~ 50 min of O$_2$ annealing, a steady state is reached in which the Pt 4f peak exhibits an unchanging shape within the timescale of the experiment. To fit the Pt 4f peak in 0.1 Torr O$_2$, we used three components at binding energies of 71.6 eV, 72.7 eV and 74.4 eV, which we assign to Pt$_{mod}$, PtO, and PtO$_2$. Here, we use the term Pt$_{mod}$ for a species with a similar peak shape to the as-deposited Pt, yet shifted by ~ 0.7 eV to higher binding energies, while PtO and PtO$_2$ are two oxidized Pt species. For details on the peak assignment, see discussion section 4.2. Over time, the ratio of the three components gradually changes in favor of the features at higher binding energies. The Pt$_{mod}$ signal decreases, whereas the PtO$_2$ signal steadily gains intensity. The PtO signal shows an initial rapid increase followed by a subsequent slow decrease. Notably, after a few minutes, the sum of the PtO and PtO$_2$ signal stays approximately constant. However, even after ~ 70 min in 0.1 Torr (~ 4x10$^8$ Langmuir), Pt$_{mod}$ remains the prominent species.

Apart from its effect on Pt 4f peak shape and position, $O_2$ annealing also induces losses in the signal intensity, which are pressure-dependent. In $1\times10^{-6}$ Torr $O_2$ at 800 K (Figure 2c), we find a signal loss of 8% over ~ 50 min (~ $3\times10^3$ Langmuir $O_2$) compared to the measurement in UHV (in the following referred to as original intensity $I_0^{UHV, 800\,K}$). Further increasing the pressure to $1\times10^{-5}$ Torr $O_2$ accelerates the Pt 4f intensity loss (Figure 2c, Figure S7) and over ~ 40 min (~ $3\times10^4$ Langmuir $O_2$), the Pt 4f signal decreases to ~ 10% of the original intensity. A subsequent return to UHV conditions stops the decrease in Pt 4f intensity but fails to restore the original state. We find that the signal loss decreases linearly with the $O_2$ dosage at a slope of ~ $4\times10^{-5}$ Langmuir$^{-1}$ both in $1\times10^{-6}$ Torr and $1\times10^{-5}$ Torr.

In 0.1 Torr $O_2$ at 800 K, the Pt 4f peak amplitude decreases over ~ 70 min (~ $4\times10^8$ Langmuir $O_2$) until reaching ~ 60 % of its original intensity. This signal loss initially occurs with a slope of ~ $-5\times10^{-9}$ Langmuir$^{-1}$, but slows down after ~ 10 min of $O_2$ exposure to ~ $-6\times10^{-10}$ Langmuir$^{-1}$ (Figure 2d). By plotting the $O_2$ dosage on a log scale, we can directly compare the development of the Pt 4f intensity in the three different investigated $O_2$ pressure regimes, as shown in Figure 2e. Even though the $O_2$ dosage at 0.1 Torr is four orders of magnitude larger than that at $1\times10^{-5}$ Torr, the rate of Pt signal loss is significantly slower at the higher pressure.

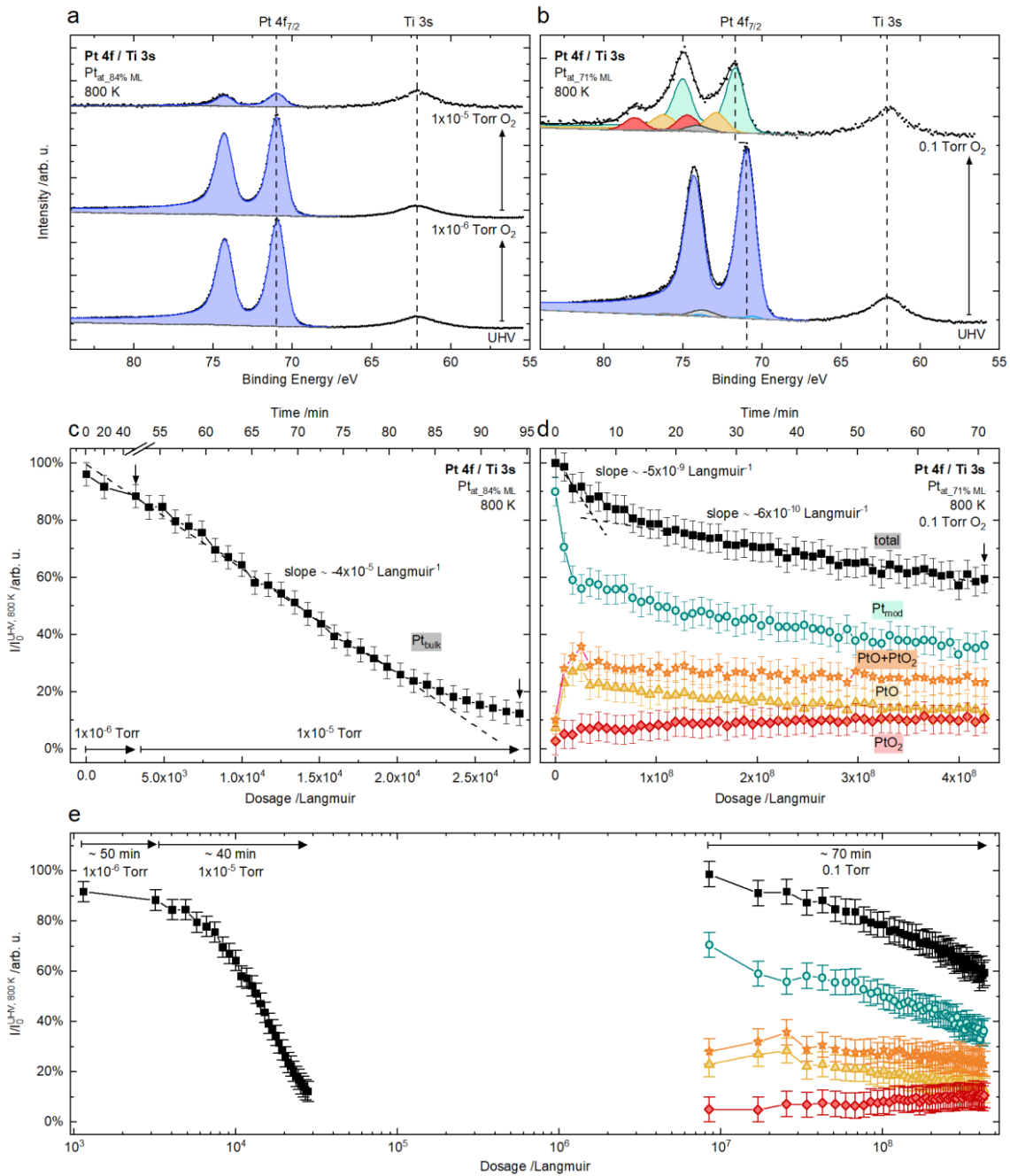

**Figure 2.** Pt 4f / Ti 3s spectra and Pt 4f peak areas for $Pt_{at\_84\% \, ML}$ (a, c, e (left)) and $Pt_{at\_71\% \, ML}$ (b, d, e (right)) in different $O_2$ pressures at 800 K recorded with an incident beam energy of 620 eV. Data are shown as black dots, fitted peak shapes as solid lines. $Pt_{at\_84\% \, ML}$ was investigated in UHV, $1\times10^{-6}$ Torr $O_2$ and $1\times10^{-5}$ Torr $O_2$. $Pt_{at\_71\% \, ML}$ was first checked in UHV and then directly exposed to 0.1 Torr $O_2$. For spectra obtained in UHV and at low $O_2$ pressures (a), a Pt bulk core level at 70.9 eV (blue) is observed. At 0.1 Torr $O_2$, three new components $Pt_{mod}$, PtO, and $PtO_2$ are found at 71.6 eV (green), 72.7 eV (yellow), and 74.4 eV (red), respectively. Additionally,

the Ti 3s plasmon (light gray) was fitted based on intensity of the Ti 3s peak. For Pt$_{at\_71\% ML}$, the Al impurity was explicitly considered (dark gray, see experimental section). Dashed lines indicate the position of the Pt 4f$_{7/2}$ (bulk and Pt$_{mod}$) and Ti 3s peak maxima. Spectra are offset for clarity. Pt 4f peak areas were obtained from the fits and normalized to the respective original intensity in UHV at 800K ($I_0^{UHV,\ 800\ K}$). The error bars were estimated based on the standard deviation of the Pt 4f peak area of four measurements at UHV and room temperature at different positions. The slope of the signal loss was determined by a linear regression analysis of the acquired total Pt 4f peak intensity. The arrows in Figure 2 c, d, and e indicate the beginning and the end of the measurement at the respective pressure.

*3.3 The Pt 4f core level peak during UHV/O$_2$ redox cycling*

According to Beck et al., the effects of annealing in high O$_2$ pressures on the Pt/TiO$_2$ system are reversible by annealing in H$_2$.[21] In order to investigate the reversibility of O$_2$ annealing for our system, we performed a series of reduction experiments (here using the inherent reducing properties of UHV) subsequent to O$_2$ exposure.

Figure 3 shows the evolution of Pt 4f / Ti 3s spectra for Pt$_{at\_6\%\,ML}$/TiO$_2$ (a, b) and Pt$_{10\_18\%\,ML}$/TiO$_2$ (c, d) during two UHV/O$_2$ annealing cycles. A third cycle was performed for Pt$_{10\_18\%\,ML}$, which is shown in Figure S9. Under the same conditions, Pt species with almost identical binding energies as for Pt$_{at\_71\%\,ML}$ (Fig. 2) are observed for Pt$_{at\_6\%\,ML}$ and Pt$_{10\_18\%\,ML}$ despite the lower metal content as well as for Pt$_{at\_84\%\,ML}$ after the exposure to O$_2$ at 1x10$^{-6}$ Torr and 1x10$^{-5}$ Torr (Figure S10). Small differences in binding energy may likely be caused by the inaccuracy of the fit at low signal/noise ratios. The ratio of Pt$_{mod}$, PtO, and PtO$_2$ differs depending on the method of deposition. For Pt$_{10\_18\%\,ML}$, PtO$_2$ exhibits the highest intensity, while for Pt$_{at}$ samples, Pt$_{mod}$ accounts for the majority of the Pt 4f peak area. High temperature reduction in UHV after annealing in elevated O$_2$ pressures (both at 800 K) restores the original Pt 4f peak shape that is characteristic for bulk Pt (for details on peak shape and positions see section 3.2). Consecutive O$_2$ annealing steps again result in the formation of up to three species, Pt$_{mod}$, PtO, and PtO$_2$. At low Pt 4f intensities, the PtO species could not be accurately resolved. Notably, the binding energies determined for Pt 4f core levels in UHV at 800 K decrease during cycling for both Pt$_{at\_6\%\,ML}$ (71.2 eV to 70.9 eV after the first cycle) and Pt$_{10\_18\%\,ML}$ (70.9 eV to 70.8 eV and 70.7 eV after the first and the second cycle, respectively).

The observed reversible changes of the Pt 4f peak shape are accompanied by an irreversible loss of the integral Pt 4f peak intensity (see section 3.2) with the first O$_2$ annealing step exhibiting the most pronounced effect (S11). Overall, our results suggest a gradual decline of the Pt 4f area during cycling, but with an accurate quantification becoming increasingly difficult as the signal intensity decreases and changes in peak area fall within the range of measurement error. In total, the Pt 4f intensity of Pt$_{at\_6\%\ ML}$ and Pt$_{10\_18\%\ ML}$ decreases to ~ 20% and ~ 4% of the original intensity, respectively (see Figure 3 and Figure S11). Note that for Pt$_{10\_18\%\ ML}$, the second and third O$_2$ annealing steps were performed with an O$_2$ pressure of 0.02 Torr and 0.2 Torr, respectively. In this elevated pressure regime, the same behavior is observed throughout.

For Pt$_{10\_18\%\ ML}$, three UHV/O$_2$ annealing cycles (see Figure 3 and Figure S11) in total reduce the Pt 4f intensity to ~ 4% of the original intensity. Note that the second and third O$_2$ annealing steps were performed with an O$_2$ pressure of 0.02 Torr and 0.2 Torr, respectively. In this elevated pressure regime, the same behavior is observed throughout.

Overall, the effects of annealing in elevated O$_2$ pressures on the Pt 4f peak shape and binding energy are reversible, but the loss of intensity is not. Likewise, UHV annealing does not restore the Pt 4f signal intensity after annealing in 1x10$^{-5}$ Torr O$_2$. Irrespective of whether the samples were exposed to O$_2$ at high or low pressures, the reductive annealing at 800 K leads to a stable sample state, i.e. a subsequent reduction treatment in 0.1 Torr H$_2$ at 800 K after the UHV anneal performed for the samples Pt$_{at\_84\%\ ML}$, Pt$_{at\_6\%\ ML}$, and Pt$_{10\_18\%\ ML}$ had no significant effect on the spectra.

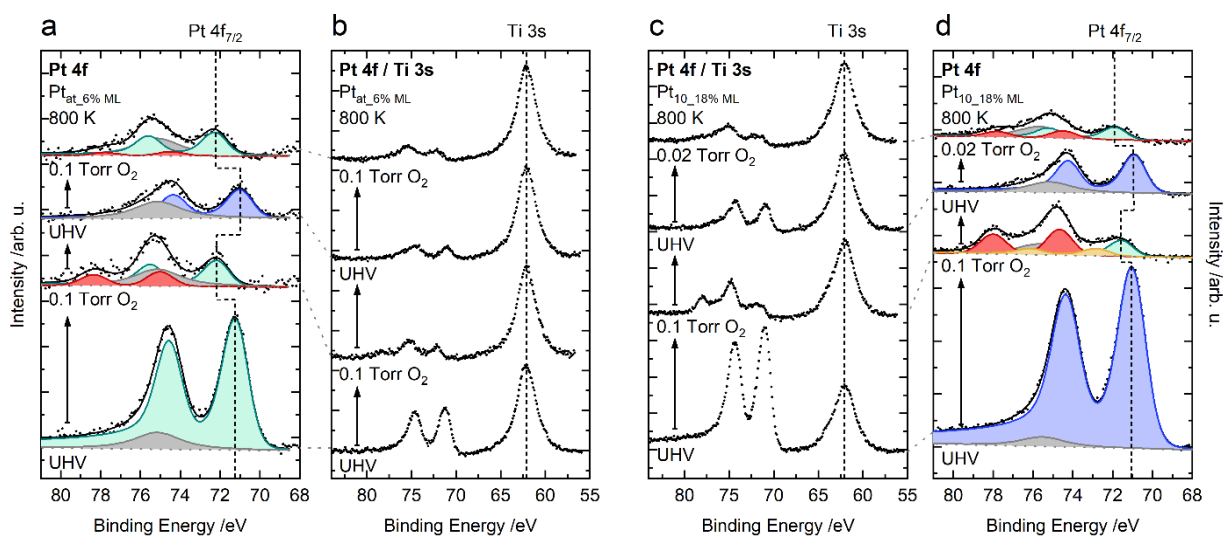

**Figure 3.** Evolution of Pt 4f / Ti 3s spectra (55-84 eV) of $Pt_{at\_6\%\ ML}$ (a, b) and $Pt_{10\_18\%\ ML}$ (c, d) upon alternating UHV and near-ambient pressure $O_2$ annealing at 800 K. The measurements were performed with an incident beam energy of 620 eV. Data are shown as black dots, fitted peak shapes as solid lines. For clarity, the same Pt 4f spectra are shown from 55-84 eV, i.e. including the Ti 3s peak (b, c), and separately from 68-81 eV, scaled to better visualize the Pt 4f fits (a,d). Under UHV conditions, a single Pt feature is observed, which was fitted based on the Pt 4f bulk core level (green and blue, respectively). At elevated $O_2$ pressures, two species $Pt_{mod}$ (green) and $PtO_2$ (red) are observed for $Pt_{at\_6\%\ ML}$ (a) and three species $Pt_{mod}$ (green), PtO (yellow), and $PtO_2$ (red) for $Pt_{10\_18\%\ ML}$ (d) (details see experimental methods section). Additionally, the Ti 3s plasmon (light gray) was fitted based on intensity of the Ti 3s peak. Dashed black lines indicate the position of the Pt $4f_{7/2}$ (bulk and $Pt_{mod}$) and Ti 3s peak maxima. Dotted gray lines indicate corresponding spectra. Spectra are offset for clarity.

*3.4 Impact of different treatments on the Ti 2p peak and correlation with the Pt 4f peaks*

After having investigated the effect of oxidizing and reducing conditions on the Pt particles by analyzing Pt 4f spectra, we now turn to the Ti 2p spectra in order to correlate potential changes in the support with the trends identified for the noble metal.

Figure 4 displays the evolution of the Ti 2p peak for $Pt_{at\_84\% ML}/TiO_2$ during oxidation and reduction (a) and the corresponding areas of the Ti $2p_{3/2}$ and Pt 4f peaks (b). The inset in Figure 4a shows the magnification of the Ti $2p_{3/2}$ low binding energy edge. All recorded Ti 2p spectra exhibit the characteristic duplet. The peak shape at room temperature and directly after reaching 800 K corresponds to that of stoichiometric $TiO_2$.[7] However, as shown in the inset of Fig. 4a, prolonged UHV annealing induces the formation of a shoulder at the low binding energy edge of the Ti $2p_{3/2}$ peak, which may be attributed to the sub-stoichiometric SMSI overlayer.[7] The shoulder persists in $1x10^{-6}$ Torr $O_2$, but vanishes completely in $1x10^{-5}$ Torr $O_2$ and is not restored by consecutive reducing treatments.

The Ti $2p_{3/2}$ area displays a sudden increase by approximately a factor of 2 compared to the original intensity upon annealing in $1x10^{-5}$ Torr $O_2$ (Figure 4b) and thus follows the opposite trend of the Pt 4f peak intensity. Note that the further increase in area observed in the consecutive measurement in UHV likely occurred predominantly during the transition from the oxygen-rich atmosphere to UHV conditions, i.e. when the sample was still exposed to $O_2$ while pumping out the chamber. Further small changes in the Ti $2p_{3/2}$ area upon exposure to different oxidizing and reducing conditions similarly reflect overall trends in the Pt 4f signal. The Ti $2p_{3/2}$ area increases slightly during the first annealing steps in UHV and $1x10^{-6}$ Torr $O_2$ and stays approximately constant after exposure to $1x10^{-5}$ Torr $O_2$. However, these changes approach the statistical error introduced when determining the Ti 2p 3/2 peak intensity and must be considered cautiously.

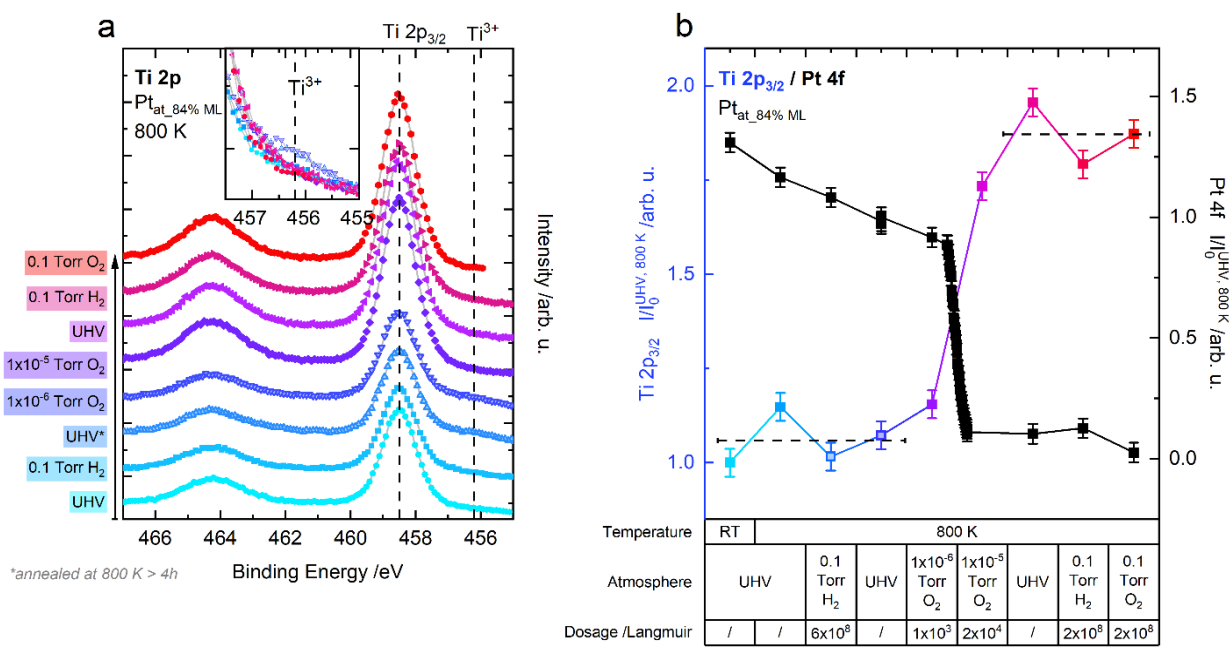

**Figure 4.** (a) Evolution of the Ti 2p spectra of the $Pt_{at\_84\%\ ML}/TiO_2$ sample in different atmospheres and temperatures. The measurements were performed with an incident beam energy of 620 eV. (a) Ti 2p spectra at 800 K measured in UHV (directly after the heating ramp), 0.1 Torr $H_2$, UHV (after >4h at 800 K), $1\times10^{-6}$ Torr $O_2$, $1\times10^{-5}$ Torr $O_2$, UHV, 0.1 Torr $H_2$, and 0.1 Torr $O_2$ (from bottom to top). These measurements correspond to the Pt 4f spectra shown in Fig. 2(a). Spectra are offset for clarity. The inset shows a magnification of the Ti $2p_{3/2}$ low binding energy edge. Dashed vertical lines indicate the position of the Ti $2p_{3/2}$ peak and of its low binding energy shoulder. Hollow shapes indicate the measurements where a Ti $2p_{3/2}$ shoulder was observed. (b) Overview of the integrated Ti $2p_{3/2}$ peak areas (455-461 eV) in different atmospheres and temperatures and corresponding Pt 4f areas (lines drawn to guide the eye). The error bars were estimated based on the standard deviation of three measurements of the Ti $2p_{3/2}$ peak area and four measurements of the Pt 4f peak, respectively, in UHV at room temperature, at different positions. Estimated Langmuir dosages are given for each pressure individually. Dashed horizontal lines indicate mean Ti $2p_{3/2}$ areas prior to and after annealing in $1\times10^{-5}$ Torr $O_2$.
20

## 4. Discussion

*4.1 Sintering and SMSI-induced encapsulation of Pt particles upon UHV annealing*

The heat treatment of Pt deposited on $TiO_2(110)$ leads to the formation, sintering, and encapsulation of Pt particles on $TiO_2(110)$. These processes are well documented[7, 9, 28, 36] and our observations are in excellent agreement with the literature. We follow the interpretation of Madey and co-workers and assign the evolution of our XPS spectra during prolonged heating to sintering and encapsulation.[7, 36] The Pt 4f core level shifts indicate sintering to particles with a metallic character, which we observe for all samples except for $Pt_{at\_6\%\,ML}$. The latter retains some characteristics of small particles, an observation that is in line with a lower Pt loading resulting in a smaller average particle size. Apart from ripening, annealing under reducing conditions (UHV or $H_2$) is further known to induce the encapsulation of Pt particles by a thin sub-stoichiometric $TiO_x$ overlayer.[6, 7, 9, 41] In accordance with XPS spectra reported by Pesty et al. for encapsulated $Pt/TiO_2(110)$ catalysts, we observe a similar decrease in the Pt 4f signal by ~ 10 % together with a decrease in its Gaussian width for all samples. Likewise, we find a Pt 4f core level shift by ~ 0.2 eV to higher binding energies for $Pt_{at\_84\%\,ML}$ upon prolonged annealing in UHV (> 4 h) and a simultaneous appearance of a shoulder at the low binding energy edge of the Ti $2p_{3/2}$ peak. These phenomena can be assigned to a covering of the Pt surface with a reduced $TiO_x$ layer based on the work of Pesty et al..[7]

*4.2 Pressure dependence of Pt oxidation and SMSI kinetics*

Our results demonstrate a strong pressure dependence in the formation of surface species of the $Pt/TiO_2$ system upon oxygen exposure at elevated temperature. In general, we differentiate between two phenomena, i.e. the oxidation of the $TiO_x$ encapsulation layer and its growth on the one hand and the oxidation of the Pt particles themselves on the other hand. We observe distinct $O_2$ pressure regimes where these two processes occur.

At intermediate $O_2$ pressures of $1\times10^{-6}$ Torr and $1\times10^{-5}$ Torr, Pt remains in the metallic state with a Pt $4f_{7/2}$ core level at 70.9 eV[35, 38, 39], while the peak linearly decreases in intensity over time. The Pt 4f signal loss is reflected in a similar increase of the Ti $2p_{3/2}$ peak intensity, suggesting an increase in the amount of $TiO_2$ close to the surface at the expense of Pt. While the support is not significantly affected by $O_2$ annealing in $1\times10^{-6}$ Torr, we observe a re-oxidation of the sub-stoichiometric SMSI overlayer in $1\times10^{-5}$ Torr, which is reflected in the decrease of the $Ti^{3+}$ shoulder in the Ti 2p spectrum. A similar oxidation was reported by Pesty et al. already in $1\times10^{-6}$ Torr. The fact that we observe the effect only at a 10 times larger oxygen pressure might be related to the higher annealing temperature of the study of Pesty et al. of up to 1000 K or to a different degree of bulk reduction of the employed $TiO_2$ support.[7]

Based on our observations, we propose that in this intermediate pressure regime, the $TiO_x$ layer encapsulating the Pt particles grows in thickness, thus leading to the gradual loss of Pt signal. In order to elucidate the mechanism of this process, it is useful to consider investigations of the growth of bare $TiO_2$ in $O_2$ and compare this process to the $Pt/TiO_2$ system. It is generally believed that at sufficiently high temperatures, $Ti^{3+}$ interstitials migrate to the crystal surface where they are oxidized and form new layers of $TiO_2$.[4, 42, 43, 44] The rate of the growth of bare $TiO_2$ has been shown to increase linearly with $O_2$ pressure.[44] Unambiguous studies on the effect of noble metal particles on this reaction are surprisingly rare. Bennett et al. investigated a $Pd/TiO_2$ system in $7.5\times10^{-8}$ Torr $O_2$ at 673 K using STM and found a de-encapsulation of Pd particles in $O_2$ and a greatly enhanced titania interstitial oxidation rate in the vicinity of Pd particles (~ 16 times faster compared to bare $TiO_2$). This growth of the support eventually resulted in the complete burial of Pd particles in the $TiO_2$ support.[45] The authors proposed a mechanism involving the adsorption and dissociation of $O_2$ on the noble metal particles followed by the spillover of $O_{ad}$ onto the support. The thus increased concentration of reactive oxygen on the $TiO_2$ surface accordingly should lead to the rapid oxidation of $Ti^{3+}$ interstitials.

This interpretation is in line with our own observations and our hypothesis of a complete burial of the Pt particles at an $O_2$ pressure of $1 \times 10^{-5}$ Torr.

Our measurements further show that in this intermediate pressure regime, the growth rate of the encapsulation layer depends linearly on the $O_2$ partial pressure for the $Pt/TiO_2$ system investigated herein. While the growth rate of bare $TiO_2$ in $O_2$ is significantly slower, its equally linear dependency on the $O_2$ partial pressure[44] suggests comparable mechanisms in both cases. The different $O_2$ pressure necessary to achieve significant oxidation of the support in our investigations compared to the work of Bennett et al. on $Pd/TiO_2$ is likely caused by a different defect concentration of the support, as this is known to strongly affect the rate of $TiO_2$ growth in $O_2$.[43, 46] While Bennett et al. sequentially observed reductive encapsulation, oxidative de-encapsulation, and a consecutive deeper oxidative burial, we do not observe evidence for the intermediate oxidative de-encapsulation.[45]

It remains an open question whether this discrepancy is due to differences in the experimental conditions, the type of noble metal, the sample preparation, or the time resolution of our experiments being insufficient to resolve this process. It is also worth stressing that most authors reporting a de-encapsulation of noble metal particles upon $O_2$ annealing employed significantly higher pressures (ambient pressure to 70 Torr) and powdered catalysts with a much higher noble metal-to-support ratio (commonly 2-5wt%).[6, 17, 18, 21, 47] We attribute the discrepancy between the full encapsulation of the noble metal described above and the incomplete burial usually observed in applied catalysis to these differences in loading and pressure, as will become clear in the comparison with our 0.1 Torr $O_2$ experiments in the following.[21, 45]

In contrast to lower $O_2$ pressures, annealing in 0.1 Torr $O_2$ at 800 K does not only affect the $TiO_x$ overlayer, but even leads to an oxidation of the Pt particles. Three distinct species at $71.6 \pm 0.02$ eV, $72.8 \pm 0.1$ eV, and $74.6 \pm 0.3$ eV (binding energy values averaged over data for $Pt_{at\_84\% ML}$, $Pt_{at\_71\% ML}$, and $Pt_{10\_18\% ML}$), can be differentiated in the Pt 4f spectra at these conditions. The binding energies of $72.8 \pm 0.1$ eV and $74.6 \pm 0.3$ eV correspond to Pt oxide species, likely

PtO and PtO$_2$.[32, 39, 48, 49, 50, 51] The feature at 71.6 eV is less easily identified and there are several possibilities which may in principle account for the core level shift. While the increase in binding energy of ~ 0.7 eV is too small to correspond to a full oxidation of Pt, similar shifts have been reported for Pt single crystals in the presence of O$_2$, where they are commonly associated with the coordination of more than one oxygen atom to Pt, possibly forming a superoxo- or peroxo-like species.[38, 39] However, for large nanoparticles or extended islands, a simple coordination of oxygen at the Pt surface without significant structural changes should lead to a coexistence of both the Pt metallic bulk core level peak at 70.9 eV and the surface PtO$_{ad}$ feature at higher binding energies, which we do not observe here.

For Pt particles, the intercalation of O$_2$ into the noble metal[52] and the formation of Pt–Ti–O alloys[21, 51] have been suggested to account for core level shifts around ~ 0.7-0.9 eV upon oxidation. While similar effects cannot be ruled out for our system, it seems unlikely that such reactions would only occur above the high O$_2$ threshold pressure necessary to facilitate the formation of the 71.6 eV species and/or in the presence of Pt oxides in our experiments. Therefore, a likely explanation is that upon Pt oxidation, small metallic domains are formed, which are situated in an oxygen-rich coordination environment (either from PtO$_x$ or TiO$_x$). Such species may likely display similarly increased binding energies as small Pt particles on TiO$_2$.[36, 37] At the same time, this coordination environment may act as a passivating layer hindering further oxidation of Pt. This hypothesis is supported by the comparable Pt 4f shifts observed prior to sintering and the decrease in the Pt oxidation rate after the first few minutes of near-ambient pressure O$_2$ exposure. We refer to the features observed after 0.1 Torr O$_2$ annealing (and after sintering for Pt$_{at\_6\% ML}$) in the range of 71.1-72.2 eV as Pt$_{mod}$, pointing out the modified nature of the Pt particles which might be similar to that of the small particles formed during Pt evaporation at room temperature based on the comparable Pt 4f core level binding energy (see Fig. 1). Reasons for the observed core level shift of Pt$_{mod}$ may include less efficient screening and/or neutralization of the photohole as well as changes in the electronic structure compared to larger metallic Pt particles. For a concise overview, the binding energies of metallic bulk Pt, $Pt_{mod}$, PtO, and $PtO_2$ found herein and the reference values from the literature, which serve as a basis for our peak assignment, are summarized in Table 1.

**Table 1.** Summary of Pt $4f_{7/2}$ binding energies for the Pt bulk core level (metallic), PtO, and $PtO_2$ in the present work and reported in the literature. Note that several different Pt species found in the presence of $O_2$ are listed under $Pt_{mod}$ together with the species assigned by us to the feature at 71.1-72.2 eV in this work.

| Sample | Pt bulk /eV | $Pt_{mod}$ /eV | PtO /eV | $PtO_2$ /eV | Ref |
|---|---|---|---|---|---|
| Pt/$TiO_2$[a] | 70.9±0.03 | 71.6±0.02 | 72.8±0.1 | 74.6±0.2 | this work |
| $Pt_{at\_6\% ML}$/$TiO_2$ | 71.2 | 72.2 | / | 74.7 | this work |
| Pt single crystal | 70.9 | 71.1, 71.4 | / | / | 35, 38 |
| Pt/$TiO_2$ (encapsulated) | ~ 71.5 | / | / | / | 7 |
| Pt/$TiO_2$ ($NO_2$ oxidation) | 71.5-71.6 | 71.6 | 73.0 | 74.6 | 52 |
| $PtO_2$ (bulk) | / | / | / | 74.1 | 48 |
| Pt/Ti/$SiO_2$/Si | 71.2 | / | 72.1 | / | 50 |

[a]The values given for Pt/$TiO_2$ from this work correspond to the average over $Pt_{at\_84\% ML}$, $Pt_{at\_71\% ML}$, and $Pt_{10\_18\% ML}$.

A similar oxidation behavior may also occur when a different oxidant than $O_2$ is used. Comparable Pt 4f spectra were obtained by Vovk et al. after exposing a Pt/$TiO_2$ sample to large amounts of $NO_2$ at room temperature.[52] Dosing 4500 Langmuir $NO_2$ induced a similar Pt 4f core level shift of 0.6 eV to higher binding energies for some of the metallic Pt, while the exposure of 30000 Langmuir $NO_2$ resulted in the formation of two additional compounds at 73.0 eV, and 74.6 eV. These two features were interpreted to originate from PtO and $PtO_2$ species.

Considering the close resemblance of the Pt 4f spectra reported by Vovk et al. after oxidation with $NO_2$ at room temperature and the spectra obtained in this work in 0.1 Torr $O_2$ at 800 K, it seems very likely that similar Pt species are formed. In order to achieve the formation of three coexisting Pt compounds in $O_2$ without employing harsher oxidizing agents, sufficiently high temperatures and $O_2$ pressures are essential. It is noteworthy that Beck et al. did not report the formation of fully oxidized Pt species under similar conditions as used in this work (0.75 Torr $O_2$, 873 K), but only found a decrease in Pt 4f intensity and a core level shift of ~ 0.7 eV to higher binding energies.[21] This may be related to the significantly larger Pt particles employed by these authors (diameter >10 nm, 2 wt% Pt) and thus a lower surface-to-volume ratio leading to a higher bulk contribution to the Pt 4f signal.

Bulk $PtO_2$ is known to decompose into metallic Pt and $O_2$ in UHV at elevated temperatures.[48] A similar behavior has been reported for oxidized Pt particles supported on $TiO_2$.[51-53] Based on these findings, cycling $O_2$ and UHV annealing should have a fully reversible effect on the oxidation state of the Pt particles. In line with this presumption, we herein observe the reduction of the oxidic Pt species PtO and $PtO_2$ to metallic Pt in UHV at 800 K and the reverse reaction in subsequent oxidation steps in elevated $O_2$ pressures at 800 K. However, $O_2$/UHV cycling at 800 K also leads to an irreversible decrease in the Pt 4f intensity, which can be attributed to a growth of the encapsulation layer just like we observed at lower $O_2$ pressures. However, the process is much less efficient at elevated $O_2$ pressures. Even though the $O_2$ exposure is substantially higher at 0.1 Torr than at $1 \times 10^{-5}$ Torr, the rate of the Pt signal loss, i.e. of the overlayer growth, is orders of magnitude lower at the higher pressure. The decrease in Pt 4f intensity is only governed by the vanishing of the $Pt_{mod}$ feature, while the total amount of Pt oxide stays approximately constant after ~ 10 min. Hence, we tentatively conclude that the overlayer grows predominantly over the $Pt_{mod}$ species.

In addition to the signal loss, the Pt 4f core level of small Pt particles ($Pt_{at\_6\% ML}/TiO_2$) gradually decreases in binding energy upon cycling. This suggests sintering into larger, more bulk-like particles, which is consistent with reports on the accelerating effect of $O_2$ at high temperatures on the ripening of Pt clusters on $TiO_2$.[30]

A possible interpretation for our observations may be given by considering the surface/interfacial energies of the resulting Pt species. The surface free energy of $TiO_2$ is lower than that of metallic Pt, but higher than that of α-$PtO_2$.[54, 55] Covering Pt particles by a $TiO_x$ overlayer therefore reduces the overall surface energy and is considered one of the main driving forces for the 'classical' SMSI.[7, 56] In UHV, the growth of the $TiO_x$ layer is self-limited, most likely due to its polarity.[9] At $1 \times 10^{-5}$ Torr $O_2$ and lower partial pressures, Pt particles retain their metallic character with high interfacial energies. At the same time, the overlayer is oxidized (as evidenced by the vanishing $Ti^{3+}$ signal at $1 \times 10^{-5}$ Torr $O_2$ for $Pt_{at\_84\% ML}$) and the restructuring expected to accompany such an oxidation may also allow for further growth in thickness. In this process, the metallic Pt will catalyze the oxidation of $Ti^{3+}$ interstitials and thus the growth of the $TiO_x$ layer in $O_2$, assuming a similar mechanism as suggested for $Pd/TiO_2$ by Bennett et al.[45].

Therefore, the burial of the noble metal by a freshly grown $TiO_2$ film is facilitated at $1 \times 10^{-5}$ Torr, which is in turn reflected in a rapid decrease of the Pt 4f peak and increase of the Ti 2p peak. We assign the driving force for this process to a reduction of the interfacial energy in the $Pt/TiO_2$ system, which occurs when the film is growing in thickness.

As a consequence of this mechanism, one would expect an increase in the growth rate of the encapsulating layer with the $O_2$ pressure, as it is observed at intermediate pressures. However, at near-ambient pressures the decrease in reaction rate suggests a more complex reaction under these conditions, which is corroborated by the different Pt species evolving. Their formation is most likely the reason for the stabilization of the system as $PtO_x$ formed at elevated $O_2$ pressures evidently catalyzes the $TiO_2$ growth much less efficiently than metallic Pt. More specifically, the loss of Pt is most rapid in the first ~ 10 min at 0.1 Torr $O_2$ while the PtO and $PtO_2$ species are still growing in intensity. As the rate of Pt oxidation slows down, the rate of the entire Pt signal loss similarly declines. DFT calculations show that it is energetically favorable for $PtO_2$ to wet the $TiO_2$ surface.[54] Hence, there is no driving force for a deeper encapsulation of $PtO_2$. This is supported by the approximately constant $PtO_x$ area after a few minutes of $O_2$ exposure at 0.1 Torr.

A lower activity of $PtO_x$ species in the oxidation of $Ti^{3+}$ is also in accordance with the lower activity of $PtO_2$ in the NO oxidation in diesel oxidation catalysts.[57] The two proposed $TiO_2$ growth mechanisms in the presence of metallic and oxidic Pt, i.e. at $1 \times 10^{-5}$ Torr and 0.1 Torr $O_2$, are illustrated in Figure 5, respectively.

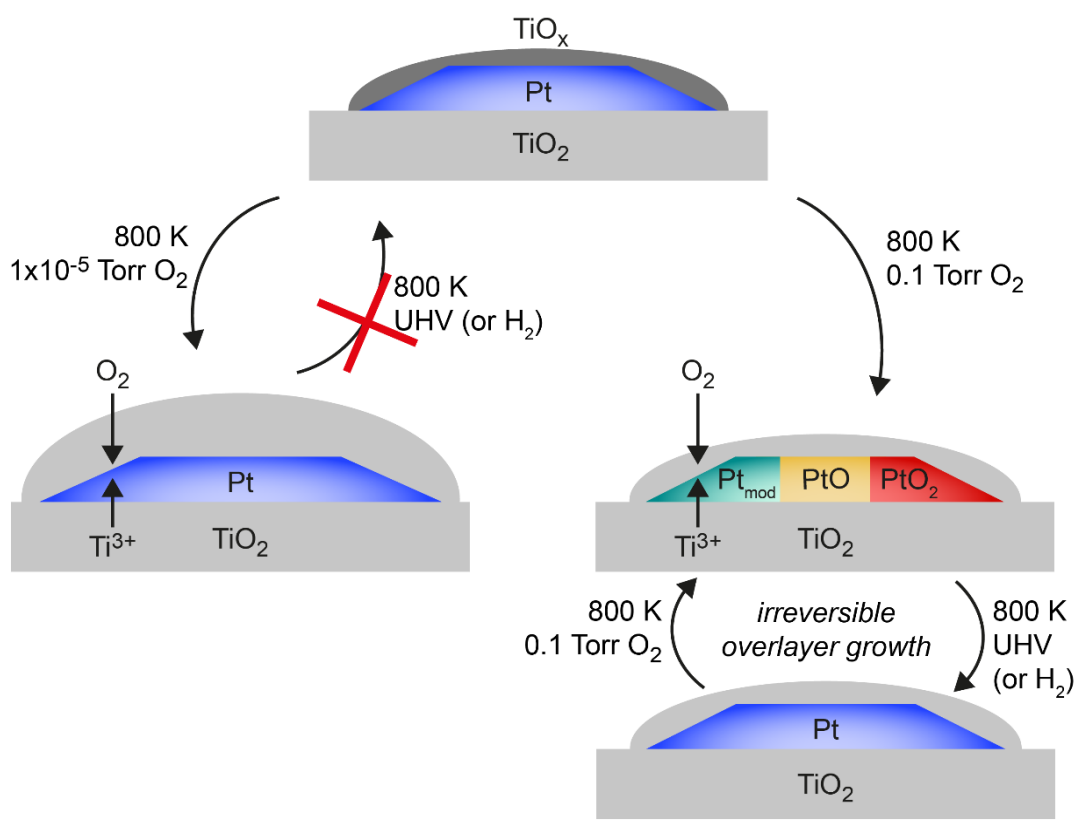

**Figure 5.** Illustration of the two proposed mechanisms for the growth of $TiO_2$ in the presence of metallic Pt (left; $1 \times 10^{-5}$ Torr $O_2$) and of $PtO_x$ (right; 0.1 Torr $O_2$) at 800 K. Metallic Pt

catalyzes the growth of the encapsulating layer at lower pressures *via* the dissociation and spillover of oxygen onto the support. At higher pressures, the chemical potential is sufficient to oxidize Pt, which does not act as a catalyst for the oxidation of the support. Due to the low surface energy of $PtO_2$, there is no driving force for an encapsulation of the oxide. The oxidation of Pt is reversible by reduction in UHV, but the growth of the encapsulating layer is irreversible, regardless of the applied $O_2$ pressure. Reduction in 0.1 Torr $H_2$ at 800 K has the same effects as UHV annealing.

## 5. Conclusion

In this work, we have investigated the effects of $O_2$ oxidation and UHV/$H_2$ reduction on well-defined Pt/$TiO_2$(110) catalysts at pressures of up to 0.1 Torr using NAP-XPS. We were able to identify three different $O_2$ pressure regimes at 800 K, which differ significantly with respect to their impact on the Pt/$TiO_2$ system. At intermediate $O_2$ pressures ($1 \times 10^{-6}$ and $1 \times 10^{-5}$ Torr), the Pt particles remain in their metallic state, but the support is oxidized and grows in thickness. The growth rate depends linearly on the pressure resulting in a rapid loss of the Pt signal intensity at $1 \times 10^{-5}$ Torr $O_2$. We propose that metallic Pt acts as a catalyst for the oxidation of $Ti^{3+}$ interstitials *via* the dissociative adsorption of $O_2$ on the metal followed by the spillover of reactive oxygen onto the support, as suggested by Bennett et al. for Pd/$TiO_2$.[45] The thus accelerated growth of the $TiO_2$ support around and over the Pt particles is evident in the rapid decrease of the Pt 4f signal intensity and a simultaneous increase in the Ti 2p signal. At an even higher pressure of 0.1 Torr $O_2$, not only the support but also the Pt particles are partially oxidized. We assign the resulting Pt 4f peaks at higher binding energies to PtO and $PtO_2$. Surprisingly, the encapsulation of Pt by the support occurs distinctly less effective at this pressure compared to $1 \times 10^{-5}$ Torr $O_2$. Hence, we conclude that the oxidic Pt species do not catalyze the oxidation of the support as efficiently as metallic Pt. While the oxidation of Pt is reversible by a reduction in UHV, the growth of the titania overlayer is not. As a result, a cycling of UHV and $O_2$ annealing steps leads to a gradual loss of the Pt signal intensity; the metallic Pt restored by UHV annealing likely induces further oxidation of the support at the beginning of every $O_2$ annealing step before Pt oxide species are formed. These findings are independent of whether the sample was prepared *via* metal vapor deposition of atoms or the deposition of size-selected clusters since strong sintering into larger nanoparticles occurred in both cases. We could further identify that the initial thin encapsulation layer is sub-stoichiometric $TiO_x$, but becomes stoichiometric as it grows into a thicker $TiO_2$ film.

In general, the present results demonstrate the importance of the $O_2$ partial pressure on platinum-decorated titania catalysts at elevated temperatures. Our study suggests that the observed processes are a result of the dynamic, pressure-dependent switching between the oxidation of the support and the Pt particles, which is of significant interest for a variety of different catalytic applications as for example the hydrogen evolution or oxidation reactions.[13, 57, 58] Our findings further highlight the importance of studies close to ambient pressure conditions for a more comprehensive understanding of the state of a catalyst under reaction conditions.

**Supporting Information**: Additional XPS spectra (surveys, Pt 4f / Ti 3s, Ti 2p, C 1s) and an overview of optimized fit parameters.

**Corresponding Author**

* Ueli Heiz, E-mail: ulrich.heiz@mytum.de

* Barbara A. J. Lechner, E-mail: bajlechner@tum.de

**Author Contributions**

The manuscript was written through contributions of all authors. All authors have given approval to the final version of the manuscript. ‡These authors contributed equally.

**Notes**

The authors declare not competing financial interest.

**Acknowledgment**

This work was funded by the Deutsche Forschungsgemeinschaft (DFG, German Research Foundation) under Germany's Excellence Strategy EXC 2089/1-390776260 and through the project CRC1441 (project number 426888090), as well as by the European Research Council (ERC) under the European Union's Horizon 2020 research and innovation program (grant agreement no. 850764). It uses resources of the Advanced Light Source, a user Facility supported by the Office of Science of the U.S. DOE under Contract DE-AC02-05CH11231. B.A.J.L. gratefully acknowledges financial support from the Young Academy of Sciences and Humanities. P.P. gratefully acknowledges financial support from the Kekulé Scholarship of the Fonds der Chemischen Industrie. M.B. was partially supported by the Condensed Phase and Interfacial Molecular Science Program in the Chemical Sciences Geosciences and Biosciences Division of the Office of Basic Energy Sciences of the U.S. Department of Energy under Contract No. DE-AC02-05CH11231.

**Abbreviations**

UHV, ultra-high vacuum; ML, monolayer, NAP-XPS, near-ambient pressure X-ray photoelectron spectroscopy; SMSI, strong metal-support interaction.

## 6. References

(1) Hadjiivanov, K. I.; Klissurski, D. G. Surface chemistry of titania (anatase) and titania-supported catalysts. *Chemical Society Reviews* **1996**, *25* (1), 61-69, 10.1039/CS9962500061. DOI: 10.1039/CS9962500061.

(2) Liu, R.; Zheng, Z.; Spurgeon, J.; Yang, X. Enhanced photoelectrochemical water-splitting performance of semiconductors by surface passivation layers. *Energy & Environmental Science* **2014**, *7* (8), 2504-2517, 10.1039/C4EE00450G. DOI: 10.1039/C4EE00450G. Reed, P. J.; Mehrabi, H.; Schichtl, Z. G.; Coridan, R. H. Enhanced Electrochemical Stability of $TiO_2$-Protected, Al-doped ZnO Transparent Conducting Oxide Synthesized by Atomic Layer Deposition. *ACS Applied Materials & Interfaces* **2018**, *10* (50), 43691-43698. DOI: 10.1021/acsami.8b16531.

(3) Akira, F.; Kenichi, H. Electrochemical Evidence for the Mechanism of the Primary Stage of Photosynthesis. *Bulletin of the Chemical Society of Japan* **1971**, *44* (4), 1148-1150. DOI: 10.1246/bcsj.44.1148. Schneider, J.; Matsuoka, M.; Takeuchi, M.; Zhang, J.; Horiuchi, Y.; Anpo, M.; Bahnemann, D. W. Understanding $TiO_2$ Photocatalysis: Mechanisms and Materials. *Chemical Reviews* **2014**, *114* (19), 9919-9986. DOI: 10.1021/cr5001892.

(4) Diebold, U. The surface science of titanium dioxide. *Surface Science Reports* **2003**, *48* (5), 53-229. DOI: https://doi.org/10.1016/S0167-5729(02)00100-0.

(5) Henderson, M. A. A surface science perspective on $TiO_2$ photocatalysis. *Surface Science Reports* **2011**, *66* (6), 185-297. DOI: https://doi.org/10.1016/j.surfrep.2011.01.001. Walenta, C. A.; Tschurl, M.; Heiz, U. Introducing catalysis in photocatalysis: What can be understood from surface science studies of alcohol photoreforming on $TiO_2$. *Journal of Physics: Condensed Matter* **2019**, *31* (47), 473002. DOI: 10.1088/1361-648x/ab351a.

(6) Tauster, S. J.; Fung, S. C.; Garten, R. L. Strong metal-support interactions. Group 8 noble metals supported on titanium dioxide. *Journal of the American Chemical Society* **1978**, *100* (1), 170-175. DOI: 10.1021/ja00469a029.

(7) Pesty, F.; Steinrück, H.-P.; Madey, T. E. Thermal stability of Pt films on $TiO_2$(110): evidence for encapsulation. *Surface Science* **1995**, *339* (1), 83-95. DOI: https://doi.org/10.1016/0039-6028(95)00605-2.

(8) Zhang, S.; Plessow, P. N.; Willis, J. J.; Dai, S.; Xu, M.; Graham, G. W.; Cargnello, M.; Abild-Pedersen, F.; Pan, X. Dynamical Observation and Detailed Description of Catalysts under Strong Metal-Support Interaction. *Nano Letters* **2016**, *16* (7), 4528-4534. DOI: 10.1021/acs.nanolett.6b01769.

(9) Dulub, O.; Hebenstreit, W.; Diebold, U. Imaging Cluster Surfaces with Atomic Resolution: The Strong Metal-Support Interaction State of Pt Supported on $TiO_2$(110). *Physical Review Letters* **2000**, *84* (16), 3646-3649. DOI: 10.1103/PhysRevLett.84.3646.

(10) Vannice, M. A.; Twu, C. C.; Moon, S. H. SMSI effects on CO adsorption and hydrogenation on Pt catalysts: I. Infrared spectra of adsorbed CO prior to and during reaction conditions. *Journal of Catalysis* **1983**, *79* (1), 70-80. DOI: https://doi.org/10.1016/0021-9517(83)90290-7. Vannice, M. A.; Twu, C. C. SMSI effects on CO adsorption and hydrogenation on Pt catalysts: Part II. Influence of support and crystallite size on the kinetics of methanation. *Journal of Catalysis* **1983**, *82* (1), 213-222. DOI: https://doi.org/10.1016/0021-9517(83)90131-8. Sen, B.; Vannice, M. A. Metal-support effects on acetone hydrogenation over platinum catalysts. *Journal of Catalysis* **1988**, *113* (1), 52-71. DOI: https://doi.org/10.1016/0021-9517(88)90237-0.

(11) Otor, H. O.; Steiner, J. B.; García-Sancho, C.; Alba-Rubio, A. C. Encapsulation Methods for Control of Catalyst Deactivation: A Review. *ACS Catalysis* **2020**, *10* (14), 7630-7656. DOI: 10.1021/acscatal.0c01569.

(12) Vannice, M. A. The Catalytic Synthesis of Hydrocarbons from Carbon Monoxide and Hydrogen. *Catalysis Reviews* **1976**, *14* (1), 153-191. DOI: 10.1080/03602457608073410. Dandekar, A.; Vannice, M. A. Crotonaldehyde Hydrogenation on Pt/TiO$_2$ and Ni/TiO$_2$ SMSI Catalysts. *Journal of Catalysis* **1999**, *183* (2), 344-354. DOI: https://doi.org/10.1006/jcat.1999.2419. Corma, A.; Serna, P.; Concepción, P.; Calvino, J. J. Transforming Nonselective into Chemoselective Metal Catalysts for the Hydrogenation of Substituted Nitroaromatics. *Journal of the American Chemical Society* **2008**, *130* (27), 8748-8753. DOI: 10.1021/ja800959g. Macino, M.; Barnes, A. J.; Althahban, S. M.; Qu, R.; Gibson, E. K.; Morgan, D. J.; Freakley, S. J.; Dimitratos, N.; Kiely, C. J.; Gao, X.; et al. Tuning of catalytic sites in Pt/TiO$_2$ catalysts for the chemoselective hydrogenation of 3-nitrostyrene. *Nature Catalysis* **2019**, *2* (10), 873-881. DOI: 10.1038/s41929-019-0334-3.
(13) Stühmeier, B. M.; Selve, S.; Patel, M. U. M.; Geppert, T. N.; Gasteiger, H. A.; El-Sayed, H. A. Highly Selective Pt/TiO$_x$ Catalysts for the Hydrogen Oxidation Reaction. *ACS Applied Energy Materials* **2019**, *2* (8), 5534-5539. DOI: 10.1021/acsaem.9b00718. Geppert, T. N.; Bosund, M.; Putkonen, M.; Stühmeier, B. M.; Pasanen, A. T.; Heikkilä, P.; Gasteiger, H. A.; El-Sayed, H. A. HOR Activity of Pt-TiO$_{2-Y}$ at Unconventionally High Potentials Explained: The Influence of SMSI on the Electrochemical Behavior of Pt. *Journal of The Electrochemical Society* **2020**, *167* (8), 084517. DOI: 10.1149/1945-7111/ab90ae.
(14) Zhang, J.; Ma, J.; Choksi, T. S.; Zhou, D.; Han, S.; Liao, Y.-F.; Yang, H. B.; Liu, D.; Zeng, Z.; Liu, W.; et al. Strong Metal–Support Interaction Boosts Activity, Selectivity, and Stability in Electrosynthesis of H$_2$O$_2$. *Journal of the American Chemical Society* **2022**, *144* (5), 2255-2263. DOI: 10.1021/jacs.1c12157.
(15) Hao, H.; Jin, B.; Liu, W.; Wu, X.; Yin, F.; Liu, S. Robust Pt@TiO$_x$/TiO$_2$ Catalysts for Hydrocarbon Combustion: Effects of Pt-TiO$_x$ Interaction and Sulfates. *ACS Catalysis* **2020**, *10* (22), 13543-13548. DOI: 10.1021/acscatal.0c03984.
(16) Bonanni, S.; Aït-Mansour, K.; Brune, H.; Harbich, W. Overcoming the Strong Metal−Support Interaction State: CO Oxidation on TiO$_2$(110)-Supported Pt Nanoclusters. *ACS Catalysis* **2011**, *1* (4), 385-389. DOI: 10.1021/cs200001y.
(17) Braunschweig, E. J.; Logan, A. D.; Datye, A. K.; Smith, D. J. Reversibility of strong metal-support interactions on Rh/TiO$_2$. *Journal of Catalysis* **1989**, *118* (1), 227-237. DOI: https://doi.org/10.1016/0021-9517(89)90313-8. Anderson, J. B. F.; Burch, R.; Cairns, J. A. The reversibility of strong metal-support interactions. A comparison of Pt/TiO$_2$ and Rh/TiO$_2$ catalysts. *Applied Catalysis* **1986**, *25* (1), 173-180. DOI: https://doi.org/10.1016/S0166-9834(00)81234-8.
(18) Herrmann, J. M.; Gravelle-Rumeau-Maillot, M.; Gravelle, P. C. A microcalorimetric study of metal-support interaction in the Pt/TiO$_2$ system. *Journal of Catalysis* **1987**, *104* (1), 136-146. DOI: https://doi.org/10.1016/0021-9517(87)90343-5.
(19) Dwyer, D. J.; Robbins, J. L.; Cameron, S. D.; Dudash, N.; Hardenbergh, J. Chemisorption and Catalysis over TiO$_2$-Modified Pt Surfaces. In *Strong Metal-Support Interactions*, ACS Symposium Series, Vol. 298; American Chemical Society, 1986; pp 21-33.
(20) Baker, R. T. K.; Prestridge, E. B.; Garten, R. L. Electron microscopy of supported metal particles II. Further studies of Pt/TiO$_2$. *Journal of Catalysis* **1979**, *59* (2), 293-302. DOI: https://doi.org/10.1016/S0021-9517(79)80033-0.
(21) Beck, A.; Huang, X.; Artiglia, L.; Zabilskiy, M.; Wang, X.; Rzepka, P.; Palagin, D.; Willinger, M.-G.; van Bokhoven, J. A. The dynamics of overlayer formation on catalyst nanoparticles and strong metal-support interaction. *Nature Communications* **2020**, *11* (1), 3220. DOI: 10.1038/s41467-020-17070-2.
(22) Naitabdi, A.; Fagiewicz, R.; Boucly, A.; Olivieri, G.; Bournel, F.; Tissot, H.; Xu, Y.; Benbalagh, R.; Silly, M. G.; Sirotti, F.; et al. Oxidation of Small Supported Platinum-based Nanoparticles Under Near-Ambient Pressure Exposure to Oxygen. *Topics in Catalysis* **2016**, *59* (5), 550-563. DOI: 10.1007/s11244-015-0529-z.

(23) Frey, H.; Beck, A.; Huang, X.; van Bokhoven, J. A.; Willinger, M. G. Dynamic interplay between metal nanoparticles and oxide support under redox conditions. *Science* **2022**, *376* (6596), 982-987. DOI: 10.1126/science.abm3371 (acccessed 2022/08/28).
(24) Li, Y.; Zhang, Y.; Qian, K.; Huang, W. Metal–Support Interactions in Metal/Oxide Catalysts and Oxide–Metal Interactions in Oxide/Metal Inverse Catalysts. *ACS Catalysis* **2022**, *12* (2), 1268-1287. DOI: 10.1021/acscatal.1c04854.
(25) Li, Z.; Smith, R. S.; Kay, B. D.; Dohnálek, Z. Determination of Absolute Coverages for Small Aliphatic Alcohols on $TiO_2(110)$. *The Journal of Physical Chemistry C* **2011**, *115* (45), 22534-22539. DOI: 10.1021/jp208228f.
(26) Grass, M. E.; Karlsson, P. G.; Aksoy, F.; Lundqvist, M.; Wannberg, B.; Mun, B. S.; Hussain, Z.; Liu, Z. New ambient pressure photoemission endstation at Advanced Light Source beamline 9.3.2. *Review of Scientific Instruments* **2010**, *81* (5), 053106. DOI: 10.1063/1.3427218 (acccessed 2022/05/28).
(27) Courtois, C.; Eder, M.; Schnabl, K.; Walenta, C. A.; Tschurl, M.; Heiz, U. Breaking Ground: Reactions in the Photocatalytic Conversion of Tertiary Alcohols on Rutile $TiO_2(110)$. *Angewandte Chemie International Edition* **2019**, Journal Article. Henderson, M. A.; Otero-Tapiab, S.; Castrob, M. E. The chemistry of methanol on the $TiO_2(110)$ surface: the influence of vacancies and coadsorbed species. *Faraday Discussions* **1999**, *114*, 313-329, Journal Article.
(28) Rieboldt, F.; Helveg, S.; Bechstein, R.; Lammich, L.; Besenbacher, F.; Lauritsen, J. V.; Wendt, S. Formation and sintering of Pt nanoparticles on vicinal rutile $TiO_2$ surfaces. *Physical Chemistry Chemical Physics* **2014**, *16*, 21289-21299. DOI: 10.1039/C4CP02716G. Rieboldt, F.; Vilhelmsen, L. B.; Koust, S.; Lauritsen, J. V.; Helveg, S.; Lammich, L.; Besenbacher, F.; Hammer, B.; Wendt, S. Nucleation and growth of Pt nanoparticles on reduced and oxidized rutile $TiO_2$ (110). *The Journal of Chemical Physics* **2014**, *141* (21), 214702. DOI: 10.1063/1.4902249.
(29) Gan, S.; Liang, Y.; Baer, D. R.; Grant, A. W. Effects of titania surface structure on the nucleation and growth of Pt nanoclusters on rutile $TiO_2(110)$. *Surface Science* **2001**, *475* (1), 159-170. DOI: https://doi.org/10.1016/S0039-6028(00)01107-9.
(30) Bonanni, S.; Aït-Mansour, K.; Harbich, W.; Brune, H. Reaction-Induced Cluster Ripening and Initial Size-Dependent Reaction Rates for CO Oxidation on $Pt_n/TiO_2(110)$-(1×1). *Journal of the American Chemical Society* **2014**, *136* (24), 8702-8707. DOI: 10.1021/ja502867r.
(31) Heiz, U.; Vanolli, F.; Trento, L.; Schneider, W.-D. Chemical reactivity of size-selected supported clusters: An experimental setup. *Review of Scientific Instruments* **1997**, *68* (5), 1986-1994. DOI: 10.1063/1.1148113.
(32) Moulder, J. F.; Stickle, W. F.; Sobol, P. E.; Bomben, K. D. *Handbook of X-ray Photoelectron Spectroscopy*; Perkin-Elmer Corporation Physical Electronics Division, 1992.
(33) Greczynski, G.; Jensen, J.; Greene, J. E.; Petrov, I.; Hultman, L. X-ray Photoelectron Spectroscopy Analyses of the Electronic Structure of Polycrystalline $Ti_{1-x}Al_xN$ Thin Films with $0 \leq x \leq 0.96$. *Surface Science Spectra* **2014**, *21* (1), 35-49. DOI: 10.1116/11.20140506 (acccessed 2022/03/23).
(34) Hofmann, S. *Auger- and X-Ray Photoelectron Spectroscopy in Materials Science*; Springer-Verlag, 2013. Thiele, J.; Barrett, N. T.; Belkhou, R.; Guillot, C.; Koundi, H. An experimental study of the growth of Co/Pt(111) by core level photoemission spectroscopy, low-energy electron diffraction and Auger electron spectroscopy. *Journal of Physics: Condensed Matter* **1994**, *6* (27), 5025-5038. DOI: 10.1088/0953-8984/6/27/012.
(35) Zhu, J. F.; Kinne, M.; Fuhrmann, T.; Denecke, R.; Steinrück, H. P. In situ high-resolution XPS studies on adsorption of NO on Pt(111). *Surface Science* **2003**, *529* (3), 384-396. DOI: https://doi.org/10.1016/S0039-6028(03)00298-X.

(36) Steinrück, H.-P.; Pesty, F.; Zhang, L.; Madey, T. E. Ultrathin films of Pt on TiO$_2$(110): Growth and chemisorption-induced surfactant effects. *Physical Review B* **1995**, *51* (4), 2427-2439. DOI: 10.1103/PhysRevB.51.2427.
(37) Isomura, N.; Wu, X.; Hirata, H.; Watanabe, Y. Cluster size dependence of Pt core-level shifts for mass-selected Pt clusters on TiO$_2$(110) surfaces. *Journal of Vacuum Science & Technology A* **2010**, *28* (5), 1141-1144. DOI: 10.1116/1.3467033.
(38) Günther, S.; Scheibe, A.; Bluhm, H.; Haevecker, M.; Kleimenov, E.; Knop-Gericke, A.; Schlögl, R.; Imbihl, R. In Situ X-ray Photoelectron Spectroscopy of Catalytic Ammonia Oxidation over a Pt(533) Surface. *The Journal of Physical Chemistry C* **2008**, *112* (39), 15382-15393. DOI: 10.1021/jp803264v. Puglia, C.; Nilsson, A.; Hernnäs, B.; Karis, O.; Bennich, P.; Mårtensson, N. Physisorbed, chemisorbed and dissociated O$_2$ on Pt(111) studied by different core level spectroscopy methods. *Surface Science* **1995**, *342* (1), 119-133. DOI: https://doi.org/10.1016/0039-6028(95)00798-9. Wang, J. G.; Li, W. X.; Borg, M.; Gustafson, J.; Mikkelsen, A.; Pedersen, T. M.; Lundgren, E.; Weissenrieder, J.; Klikovits, J.; Schmid, M.; et al. One-Dimensional PtO$_2$ at Pt Steps: Formation and Reaction with CO. *Physical Review Letters* **2005**, *95* (25), 256102. DOI: 10.1103/PhysRevLett.95.256102.
(39) Kim, Y. S.; Bostwick, A.; Rotenberg, E.; Ross, P. N.; Hong, S. C.; Mun, B. S. The study of oxygen molecules on Pt (111) surface with high resolution x-ray photoemission spectroscopy. *The Journal of Chemical Physics* **2010**, *133* (3), 034501. DOI: 10.1063/1.3458910 (acccessed 2022/03/31).
(40) Ocal, C.; Ferrer, S. The strong metal–support interaction (SMSI) in Pt-TiO$_2$ model catalysts. A new CO adsorption state on Pt-Ti atoms. *The Journal of Chemical Physics* **1986**, *84* (11), 6474-6478. DOI: 10.1063/1.450743 (acccessed 2022/03/25).
(41) Sexton, B. A.; Hughes, A. E.; Foger, K. XPS investigation of strong metal-support interactions on Group IIIa–Va oxides. *Journal of Catalysis* **1982**, *77* (1), 85-93. DOI: https://doi.org/10.1016/0021-9517(82)90149-X.
(42) Li, M.; Hebenstreit, W.; Diebold, U. Oxygen-induced restructuring of the rutile TiO$_2$(110)(1×1) surface. *Surface Science* **1998**, *414* (1), L951 - L956. Diebold, U. Structure and properties of TiO$_2$ surfaces: a brief review. *Applied Physics A* **2003**, *76* (5), 681-687. DOI: 10.1007/s00339-002-2004-5. Krischok, S.; Günster, J.; Goodman, D. W.; Höfft, O.; Kempter, V. MIES and UPS(HeI) studies on reduced TiO$_2$(110). *Surface and Interface Analysis* **2005**, *37* (1), 77-82, https://doi.org/10.1002/sia.2013. DOI: https://doi.org/10.1002/sia.2013 (acccessed 2022/03/29). Bowker, M.; Bennett, R. A. The role of Ti$^{3+}$ interstitials in TiO$_2$(110) reduction and oxidation. *Journal of Physics: Condensed Matter* **2009**, *21* (47), 474224. DOI: 10.1088/0953-8984/21/47/474224.
(43) Li, M.; Hebenstreit, W.; Diebold, U.; A. Henderson, M.; R. Jennison, D. Oxygen-induced restructuring of rutile TiO$_2$(110): formation mechanism, atomic models, and influence on surface chemistry. *Faraday Discussions* **1999**, *114* (0), 245-258, 10.1039/A903598B. DOI: 10.1039/A903598B.
(44) Smith, R. D.; Bennett, R. A.; Bowker, M. Measurement of the surface-growth kinetics of reduced TiO$_2$(110) during reoxidation using time-resolved scanning tunneling microscopy. *Physical Review B* **2002**, *66* (3), 035409. DOI: 10.1103/PhysRevB.66.035409.
(45) Bennett, R. A.; Stone, P.; Bowker, M. Pd nanoparticle enhanced re-oxidation of non-stoichiometric TiO$_2$: STM imaging of spillover and a new form of SMSI. *Catalysis Letters* **1999**, *59* (2), 99-105. DOI: 10.1023/A:1019053512230.
(46) Li, M.; Hebenstreit, W.; Diebold, U.; Tyryshkin, A. M.; Bowman, M. K.; Dunham, G. G.; Henderson, M. A. The Influence of the Bulk Reduction State on the Surface Structure and Morphology of Rutile TiO$_2$(110) Single Crystals. *The Journal of Physical Chemistry B* **2000**, *104* (20), 4944-4950.
(47) Bernal, S.; Botana, F. J.; Calvino, J. J.; López, C.; Pérez-Omil, J. A.; Rodríguez-Izquierdo, J. M. High-resolution electron microscopy investigation of metal–support interactions in Rh/TiO$_2$. *Journal of the Chemical Society, Faraday Transactions* **1996**, *92* (15), 2799-2809, 10.1039/FT9969202799. DOI: 10.1039/FT9969202799. Penner, S.; Wang, D.; Su, D. S.; Rupprechter, G.; Podloucky, R.; Schlögl, R.; Hayek, K. Platinum nanocrystals supported by silica, alumina and ceria: metal–support interaction due to high-temperature reduction in hydrogen. *Surface Science* **2003**, *532-535*, 276-280. DOI: https://doi.org/10.1016/S0039-6028(03)00198-5.
(48) Peuckert, M.; Bonzel, H. P. Characterization of oxidized platinum surfaces by X-ray photoelectron spectroscopy. *Surface Science* **1984**, *145* (1), 239-259. DOI: https://doi.org/10.1016/0039-6028(84)90778-7.
(49) Parkinson, C. R.; Walker, M.; McConville, C. F. Reaction of atomic oxygen with a Pt(111) surface: chemical and structural determination using XPS, CAICISS and LEED. *Surface Science* **2003**, *545* (1), 19-33. DOI: https://doi.org/10.1016/j.susc.2003.08.029. Fantauzzi, D.; Krick Calderón, S.; Mueller, J. E.; Grabau, M.; Papp, C.; Steinrück, H.-P.; Senftle, T. P.; van Duin, A. C. T.; Jacob, T. Growth of Stable Surface Oxides on Pt(111) at Near-Ambient Pressures. *Angewandte Chemie International Edition* **2017**, *56* (10), 2594-2598, https://doi.org/10.1002/anie.201609317. DOI: https://doi.org/10.1002/anie.201609317 (acccessed 2022/03/31).
(50) Jung, M.-C.; Kim, H.-D.; Han, M.; Jo, W.; Kim, D. C. X-Ray Photoelectron Spectroscopy Study of Pt-Oxide Thin Films Deposited by Reactive Sputtering Using O$_2$/Ar Gas Mixtures. *Japanese Journal of Applied Physics* **1999**, *38* (Part 1, No. 8), 4872-4875. DOI: 10.1143/jjap.38.4872.
(51) Ono, L. K.; Yuan, B.; Heinrich, H.; Roldan Cuenya, B. Formation and Thermal Stability of Platinum Oxides on Size-Selected Platinum Nanoparticles: Support Effects. *The Journal of Physical Chemistry C* **2010**, *114* (50), 22119-22133. DOI: 10.1021/jp1086703.
(52) Vovk, E. I.; Kalinkin, A. V.; Smirnov, M. Y.; Klembovskii, I. O.; Bukhtiyarov, V. I. XPS Study of Stability and Reactivity of Oxidized Pt Nanoparticles Supported on TiO$_2$. *The Journal of Physical Chemistry C* **2017**, *121* (32), 17297-17304. DOI: 10.1021/acs.jpcc.7b04569.
(53) Ono, L. K.; Croy, J. R.; Heinrich, H.; Roldan Cuenya, B. Oxygen Chemisorption, Formation, and Thermal Stability of Pt Oxides on Pt Nanoparticles Supported on SiO$_2$/Si(001): Size Effects. *The Journal of Physical Chemistry C* **2011**, *115* (34), 16856-16866. DOI: 10.1021/jp204743q.
(54) Jonayat, A. S. M.; Chen, S.; van Duin, A. C. T.; Janik, M. Predicting Monolayer Oxide Stability over Low-Index Surfaces of TiO$_2$ Polymorphs Using ab Initio Thermodynamics. *Langmuir* **2018**, *34* (39), 11685-11694. DOI: 10.1021/acs.langmuir.8b02426.
(55) Vitos, L.; Ruban, A. V.; Skriver, H. L.; Kollár, J. The surface energy of metals. *Surface Science* **1998**, *411* (1), 186-202. DOI: https://doi.org/10.1016/S0039-6028(98)00363-X. Overbury, S. H.; Bertrand, P. A.; Somorjai, G. A. Surface composition of binary systems. Prediction of surface phase diagrams of solid solutions. *Chemical Reviews* **1975**, *75* (5), 547-560. DOI: 10.1021/cr60297a001.
(56) Fu, Q.; Wagner, T.; Olliges, S.; Carstanjen, H.-D. Metal−Oxide Interfacial Reactions: Encapsulation of Pd on TiO$_2$ (110). *The Journal of Physical Chemistry B* **2005**, *109* (2), 944-951. DOI: 10.1021/jp046091u. Chen, P.; Gao, Y.; Castell, M. R. Thermodynamics driving the strong metal-support interaction: Titanate encapsulation of supported Pd nanocrystals. *Physical Review Materials* **2021**, *5* (7), 075001. DOI: 10.1103/PhysRevMaterials.5.075001.
(57) Hauff, K.; Tuttlies, U.; Eigenberger, G.; Nieken, U. Platinum oxide formation and reduction during NO oxidation on a diesel oxidation catalyst – Experimental results. *Applied Catalysis B: Environmental* **2012**, *123-124*, 107-116. DOI: https://doi.org/10.1016/j.apcatb.2012.04.008.
(58) Alayon, E. M. C.; Singh, J.; Nachtegaal, M.; Harfouche, M.; van Bokhoven, J. A. On highly active partially oxidized platinum in carbon monoxide oxidation over supported platinum catalysts. *Journal of Catalysis* **2009**, *263* (2), 228-238. DOI: https://doi.org/10.1016/j.jcat.2009.02.010. Singh, J.; Nachtegaal, M.; Alayon, E. M. C.; Stötzel, J.; van Bokhoven, J. A. Dynamic Structure Changes of a Heterogeneous Catalyst within a Reactor: Oscillations in CO Oxidation over a Supported Platinum Catalyst. *ChemCatChem* **2010**, *2* (6), 653-657, https://doi.org/10.1002/cctc.201000061. DOI: https://doi.org/10.1002/cctc.201000061 (acccessed 2022/04/06). Hendriksen, B. L. M.; Frenken, J. W. M. CO Oxidation on Pt(110): Scanning Tunneling Microscopy Inside a High-Pressure Flow Reactor. *Physical Review Letters* **2002**, *89* (4), 046101. DOI: 10.1103/PhysRevLett.89.046101.

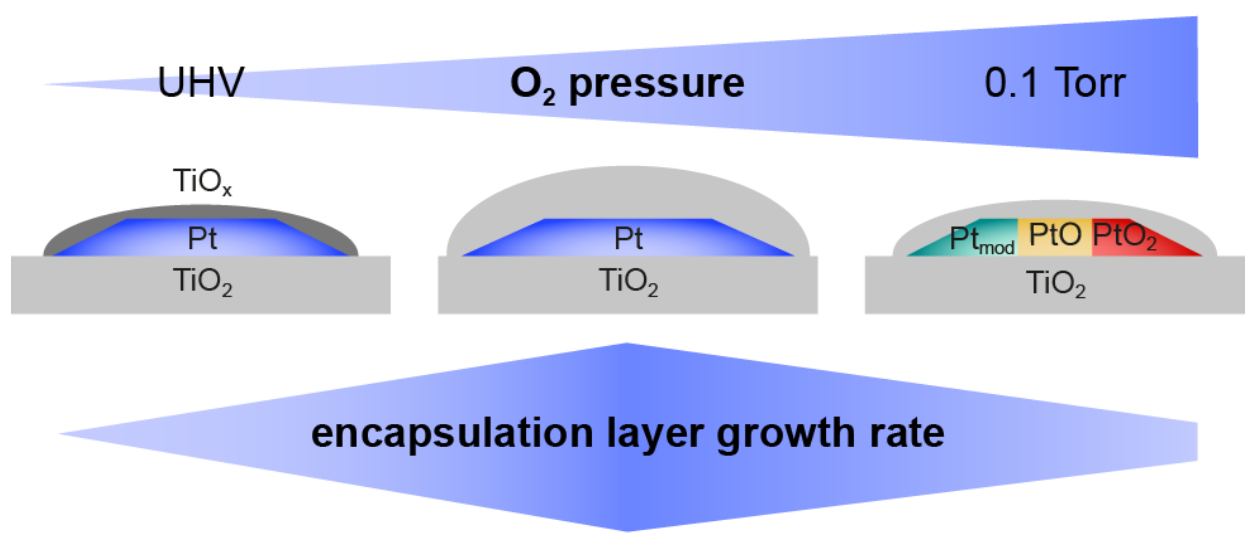